\documentclass[letterpaper,twocolumn,prl,aps,superscriptaddress,amsmath,amssymb,floatfix]{revtex4-1}
\usepackage{color}
\usepackage{amsmath}
\usepackage{amssymb}
\usepackage{graphicx}
\usepackage{esint}
\usepackage{times}

\usepackage[unicode=true,
 bookmarks=true,bookmarksnumbered=false,bookmarksopen=false,
 breaklinks=false,pdfborder={0 0 1},backref=false,colorlinks=true]
 {hyperref}
\hypersetup{
 linkcolor=magenta, urlcolor=blue, citecolor=blue, pdfstartview={FitH}, hyperfootnotes=false, unicode=true}

\makeatletter

\usepackage{textcomp}
\usepackage{epstopdf}

\pdfpageheight\paperheight
\pdfpagewidth\paperwidth

\@ifundefined{textcolor}{}{%
 \definecolor{BLACK}{gray}{0}
 \definecolor{WHITE}{gray}{1}
 \definecolor{RED}{rgb}{1,0,0}
 \definecolor{GREEN}{rgb}{0,1,0}
 \definecolor{BLUE}{rgb}{0,0,1}
 \definecolor{CYAN}{cmyk}{1,0,0,0}
 \definecolor{MAGENTA}{cmyk}{0,1,0,0}
 \definecolor{YELLOW}{cmyk}{0,0,1,0}
}

\usepackage{xcolor}\usepackage{soul}
\newcommand{\bra}[1]{\ensuremath{\left\langle#1\right|}}
\newcommand{\ket}[1]{\ensuremath{\left|#1\right\rangle}}

\definecolor{blue}{rgb}{0,0,1}
\definecolor{red}{rgb}{1,0,0}
\definecolor{green}{rgb}{0,1,0}

\makeatother

\begin{document}

\title{Quantum generative adversarial learning in a superconducting quantum circuit}

\author{L.~Hu}

\thanks{These two authors contributed equally to this work.}

\affiliation{Center for Quantum Information, Institute for Interdisciplinary Information
Sciences, Tsinghua University, Beijing 100084, China}

\author{S.~Wu}

\thanks{These two authors contributed equally to this work.}

\affiliation{Key Laboratory of Quantum Information, CAS, University of Science
and Technology of China, Hefei, Anhui 230026, P. R. China}

\author{W.~Cai}

\affiliation{Center for Quantum Information, Institute for Interdisciplinary Information
Sciences, Tsinghua University, Beijing 100084, China}

\author{Y.~Ma}

\affiliation{Center for Quantum Information, Institute for Interdisciplinary Information
Sciences, Tsinghua University, Beijing 100084, China}

\author{X.~Mu}

\affiliation{Center for Quantum Information, Institute for Interdisciplinary Information
Sciences, Tsinghua University, Beijing 100084, China}

\author{Y.~Xu}

\affiliation{Center for Quantum Information, Institute for Interdisciplinary Information
Sciences, Tsinghua University, Beijing 100084, China}

\author{H.~Wang}

\affiliation{Center for Quantum Information, Institute for Interdisciplinary Information
Sciences, Tsinghua University, Beijing 100084, China}

\author{Y.~P.~Song}

\affiliation{Center for Quantum Information, Institute for Interdisciplinary Information
Sciences, Tsinghua University, Beijing 100084, China}

\author{D.-L.~Deng}
\email{dldeng@tsinghua.edu.cn}
\affiliation{Center for Quantum Information, Institute for Interdisciplinary Information
Sciences, Tsinghua University, Beijing 100084, China}

\author{C.-L.~Zou}
\email{clzou321@ustc.edu.cn}

%\selectlanguage{english}%

\affiliation{Key Laboratory of Quantum Information, CAS, University of Science
and Technology of China, Hefei, Anhui 230026, P. R. China}

\author{L.~Sun}
\email{luyansun@tsinghua.edu.cn}

%\selectlanguage{english}%

\affiliation{Center for Quantum Information, Institute for Interdisciplinary Information
Sciences, Tsinghua University, Beijing 100084, China}

%\date{\today}
\begin{abstract}
Generative adversarial learning \cite{goodfellow2014generative} is one of the most exciting recent breakthroughs in machine learning---a subfield of artificial intelligence that is currently driving a revolution in many aspects of modern society. It has shown splendid performance in a variety of challenging tasks such as image and video generations~\cite{Creswell2018Generative}. 
More recently, a quantum version of generative adversarial learning has been theoretically proposed \cite{Lloyd2018,DallaireDemers2018} and shown to possess the potential of exhibiting an exponential advantage over its classical counterpart \cite{Lloyd2018}. Here, we report the first proof-of-principle experimental demonstration of quantum generative adversarial learning in a superconducting quantum circuit. We demonstrate that, after several rounds of adversarial learning, a quantum state generator can be trained to replicate the statistics of the quantum data output from a digital qubit channel simulator, with a high fidelity ($98.8\%$ on average) that the discriminator cannot distinguish between the true and the generated data. Our results pave the way for experimentally exploring the  intriguing long-sought-after quantum advantages in machine learning tasks with noisy intermediate-scale quantum devices \cite{Preskill2018}. 
\end{abstract}

\maketitle

Machine learning \cite{Jordan2015Machine}, or more broadly artificial intelligence~\cite{Russell2016Artificial}, represents an important area with general practical applications where near-term quantum devices may offer a significant speed-up over classical ones. With this vision, an intriguing  interdisciplinary field of quantum machine learning/artificial intelligence has emerged and attracted tremendous attentions in recent years \cite{Biamonte2017,Dunjko2018}. A  number of quantum algorithms that promise exponential speed-ups have been theoretically proposed \cite{Biamonte2017, Dunjko2018, Ciliberto2017Quantum, Gao2017anEfficient} and some were demonstrated in proof-of-principle experiments \cite{Otterbach2017,Li2015Experimental}. Yet, in most of these previous scenarios, the input data sets considered are typically classical. As a result, certain costly processes or techniques, such as quantum random access memories \cite{Giovannetti2008}, are required to first map the classical data to quantum wave-vectors so as to be processed by quantum devices, rendering the potential speed-ups nullified \cite{Aaronson2015Read}.

In this Letter, we  experimentally demonstrate a quantum version of generative adversarial network (QGAN)~\cite{Lloyd2018,DallaireDemers2018}, where both the input and output data sets are quantum from the beginning. In classical machine learning, a generative adversarial network (GAN) contains two major components, a generator (G) and a discriminator (D)~\cite{goodfellow2014generative}. They are trained through an adversarial learning procedure: in each learning round, D optimizes her strategies to identify the fake data produced by G, while G updates his strategies to fool D. Under reasonable assumptions, such an adversarial game will end up at a Nash equilibrium point, where G produces data that match the statistics of the true data, and D can no longer distinguish the fake data with a probability larger than $1/2$. In the quantum setting considered here, G consists of a superconducting circuit, which can generate an ensemble of quantum states with certain probability distribution, while D is composed of a quantum apparatus that carries out projective measurements. The arbitrary input quantum data is generated by a digital qubit channel simulator.

\begin{figure}
\includegraphics[width=1\columnwidth]{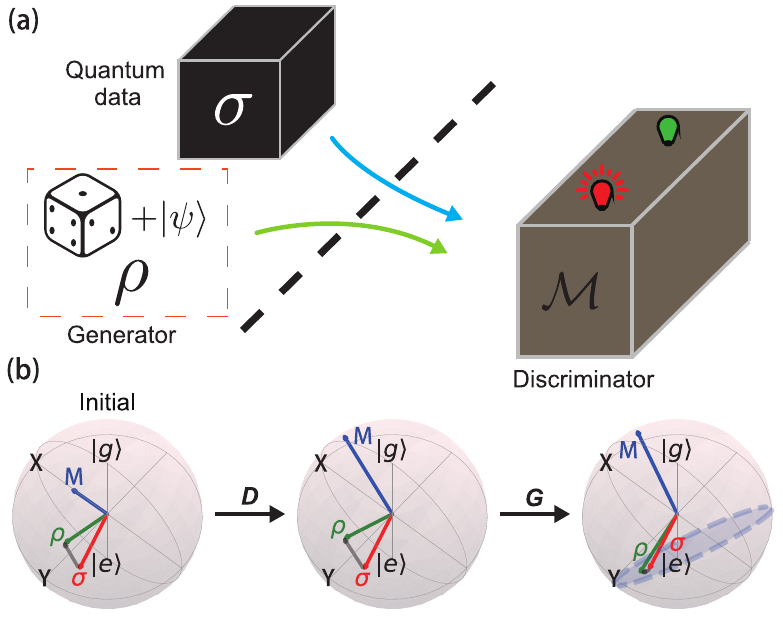} \caption{\label{fig:Fig1} \textbf{The quantum generative adversarial network
(QGAN)}. (a) The basic components of QGAN, including a black box quantum
process for the quantum true data $\sigma$, the generator (G) that
produces an ensemble of pure quantum states $\rho$, and the discriminator
(D) that performs projective measurements $\mathcal{M}$. (b) The
process of QGAN with the quantum states and the measurement basis
on a Bloch sphere, where $\left\{ \left|g\right\rangle ,\left|e\right\rangle \right\} $
are the ground and excited state of a qubit. D and G play the adversarial
game alternatively, in which D optimizes the measurement strategy
to discriminate $\rho$ from $\sigma$, while G optimizes the generation
strategy to fool D.}
\vspace{-6pt}
\end{figure}

\begin{figure*}
\includegraphics{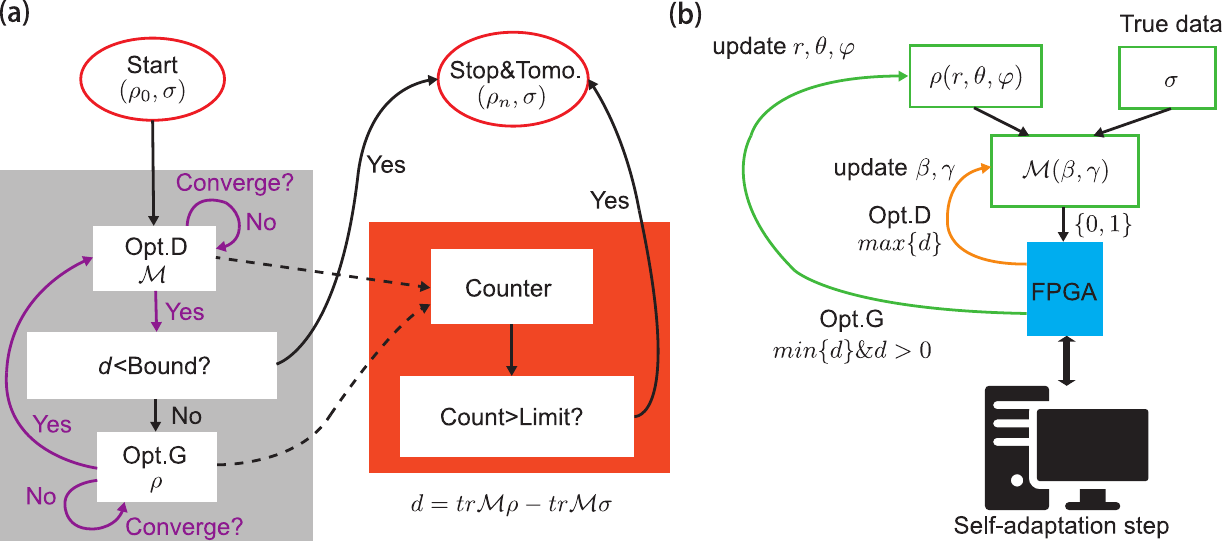} \caption{\textbf{The experimental protocol of QGAN algorithm.} \textbf{(a)}
The experiment starts with a quantum state $\sigma$ as the quantum
true data and a randomly generated state $\rho(r_{0},\theta_{0},\varphi_{0})$
from G. Then, D and G optimize their strategies to compete against
each other alternatively and repetitively. The stop condition of the
game is either D fails to distinguish $\rho$ from $\sigma$ (the
measurement output difference $d<d_{\mathrm{B}}$, a preset bound)
or the step count $c_{\mathrm{step}}$ reaches the limit $c_{\mathrm{B}}$.
\textbf{(b)} Procedure of optimizing D and G by the gradient decent
method. The initial measurement axis $\mathcal{M}\left(\beta_{0},\gamma_{0}\right)$
for D is randomly selected. The parameters $\beta$ and $\gamma$
are updated in the process of optimizing D, while $r$, $\theta$,
and $\varphi$ are updated in the process of optimizing G. The
measurement and control of the quantum system are realized through
field programmable gate arrays (FPGA), while the estimations of the
gradients are performed in a classical computer.}
%and the initial measurement axis $\mathcal{M}\left(\beta_{0},\gamma_{0}\right)$ is also randomly selected
\label{fig:Fig2} 
\end{figure*}

Figure$\,$\ref{fig:Fig1}(a) shows the schematic of the QGAN scheme.
The black box provides the quantum true data which is described by
a density matrix $\sigma$ of a quantum system, while both the internal physical
structure and the quantum process do not need to be known. G can generate
arbitrary quantum states $\left(\rho\right)$ by producing an ensemble
of pure quantum states, i.e. a pure state from a set is randomly selected
with certain probability to mimic the quantum true data. D performs quantum measurements $\left(\mathcal{M}\right)$ on the
true and the generated (fake) data, and attempts to distinguish
them by the statistics of the measurement outcomes $p_{\rho}=\mathrm{tr}\mathcal{M}\rho$
and $p_{\sigma}=\mathrm{tr}\mathcal{M}\sigma$. In the QGAN,
the measurement outcomes are public to both G and D. According to
$p_{\rho,\sigma}$, D and G compete against each other by adaptively
adjusting their strategies alternatively to distinguish $\rho$ from
$\sigma$ and to fool D, respectively. It is interesting to
note that $\sigma$ and $\rho$ represent two distinct interpretations
of mixed quantum states: one is the output of a physical process in
which an initial pure state might be entangled with some degrees of
freedom of the environment; the other is an ensemble of pure states.
Our QGAN scheme can also be explained as a game trying to distinguish
between these two interpretations.

A visualized illustration of the general procedure of QGAN is depicted
in Fig.$\,$\ref{fig:Fig1}(b) by presenting $\sigma$, $\rho$ and
$\mathcal{M}$ of a qubit system in the Bloch sphere (note that we use the same notation $\mathcal{M}$ to represent both the projective measurement and its corresponding axis). D and G play
the game alternatively. D always starts first, and in her turn $\mathcal{M}$ is optimized to maximize the difference of the measurement
outcome $d=p_{\rho}-p_{\sigma}$. In an ideal case, D's turn ends
up with $d=\frac{1}{2}\left\Vert \rho-\sigma\right\Vert _{1}$ corresponding
to the normalized trace distance~\cite{Nielsen}, and $\mathcal{M}$
converges to be parallel with $\rho-\sigma$ in the Bloch sphere
representation {[}Fig.$\,$\ref{fig:Fig1}(b){]}. For G's turn, $\rho$
is optimized to minimize $d$, and thus approaches a cross-section
such that $\rho-\sigma$ is perpendicular to $\mathcal{M}$ {[}Fig.$\,$\ref{fig:Fig1}(b){]}.
As a result, the trace distance between the fake and the true data reduces progressively in each round, and the game eventually
approaches the unique Nash equilibrium with $d=0$ and $p_{\sigma}=p_{\rho}=\frac{1}{2}$~\cite{Lloyd2018}.

\begin{figure*}
\includegraphics{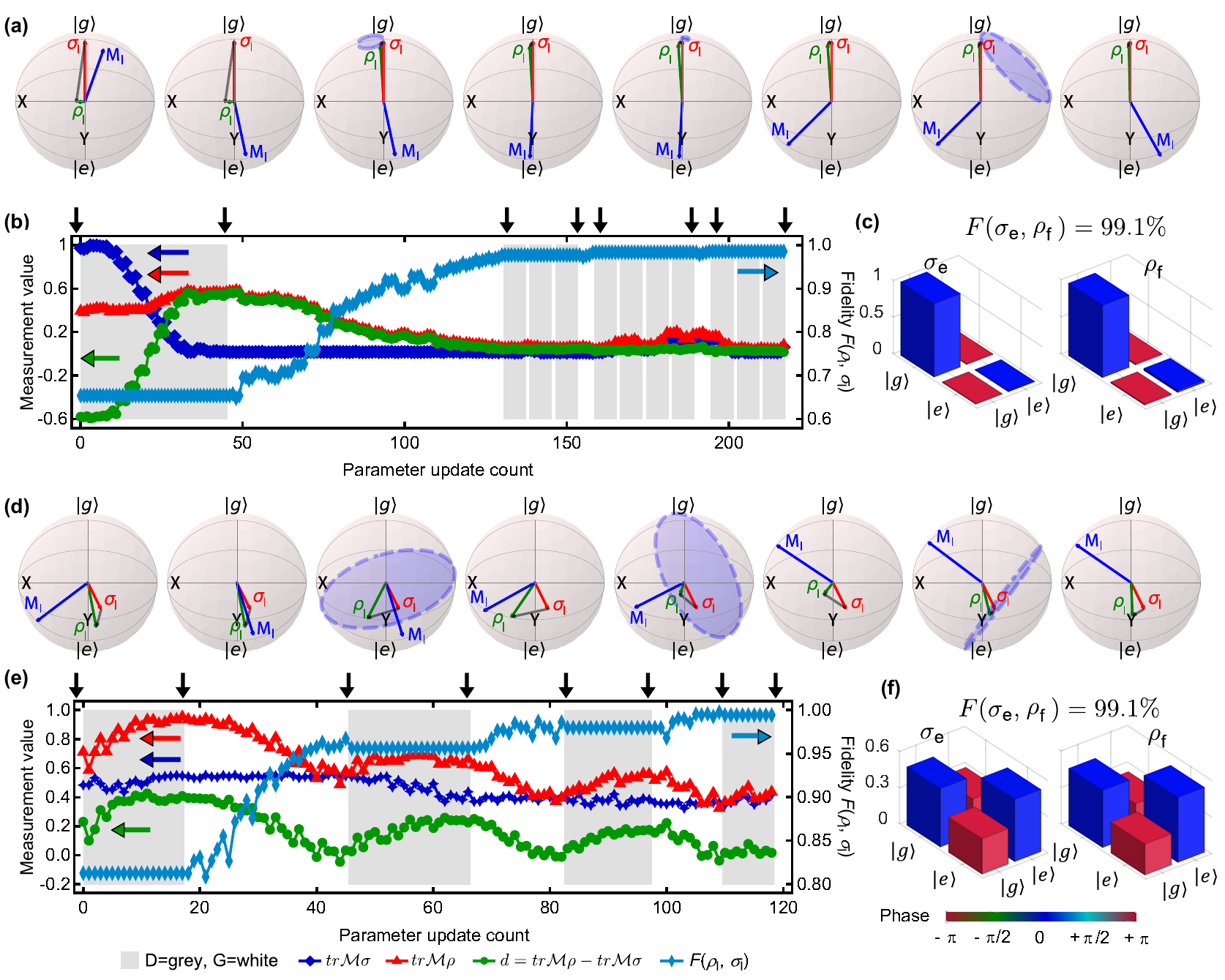}\caption{\textbf{Tracking of QGAN.} \textbf{(a-c)} Experimental results for
selecting $\sigma=\ket{g}\bra{g}$ as the quantum true data. \textbf{(a)}
shows the snapshots of the system at the particular steps indicated
by arrows in \textbf{b} (from left to right in the same order) in
the Bloch sphere representation. \textbf{(b)} shows the tracking of
$p_{\sigma}$, $p_{\rho}$, ${d}$ and $F$ during the quantum adversarial
learning process. The gray shadow regions are the processes of optimizing
D, while the rests are those for optimizing G. \textbf{(c)} shows
the measured state tomography of the experimental $\sigma_{\mathrm{e}}$
and final $\rho_{\mathrm{f}}$ with a state fidelity $F=0.991$, demonstrating
a successful QGAN that G can fool D by generating quantum data highly
similar to the true data. \textbf{(d-f)} Typical experimental results
for $\sigma$ in an arbitrary mixed state with each panel being the
counterpart of \textbf{(a-c)} respectively.}
\label{fig:Fig3} \vspace{-6pt}
\end{figure*}

We realize the QGAN learning algorithm~\cite{Lloyd2018} in a superconducting
quantum electrodynamics architecture~\cite{Devoret2013,GU2017}. Our experimental
device consists of a superconducting transmon qubit dispersively coupled
to a bosonic microwave mode~\cite{Wallraff,Paik2011}. The quantum state of the transmon qubit serves as either $\rho$ or $\sigma$ alternatively in the algorithm. The bosonic mode facilitates the creation of the quantum true data $\sigma$ in an arbitrary state through a digital quantum channel simulator, which requires adaptive control of both the transmon qubit and the bosonic mode. The detailed descriptions of the experimental device and apparatus are provided in the Supporting Materials,
and can also be found in Ref.~\cite{Hu2018b}. G generates the state $\rho(r,\theta,\varphi)$ of
the transmon by randomly preparing a pure state in the set $\left\{ U\left(\theta,\varphi\right)\left|g\right\rangle ,U\left(\pi-\theta,\varphi+\pi\right)\left|g\right\rangle \right\} $
with the corresponding probabilities $\left\{ r,1-r\right\}$.
Here, $U\left(\theta,\varphi\right)=e^{i\varphi\sigma_{\mathrm{z}}/2}e^{i\theta\sigma_{\mathrm{x}}/2}$
is the unitary operation on the transmon qubit, with $\sigma_{\mathrm{x}}$
and $\sigma_{\mathrm{z}}$ being the Pauli matrices. D performs the measurements
by applying a unitary pre-rotating operation with the axis angles ($\beta$,
$\gamma$) on the transmon and detecting the population of the ground
state $\left|g\right\rangle $, which leads to $\mathcal{M}=U^{\dagger}\left(\beta,\gamma\right)\left|g\right\rangle \left\langle g\right|U\left(\beta,\gamma\right)$.

\begin{figure*}
\includegraphics{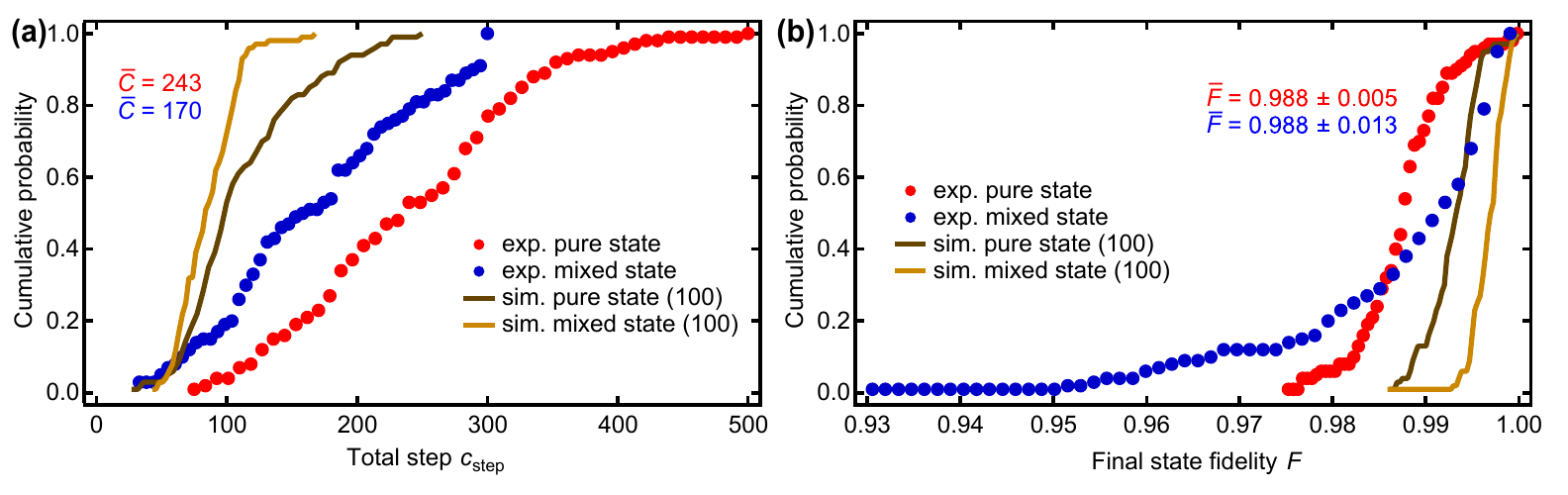} \caption{\textbf{Statistics of QGAN performance.} \textbf{(a)} The cumulative probability of the total step count to finish the adversarial learning process. The QGAN is implemented for two difference cases, with a
pure ($\ket{g}\bra{g}$) and an arbitrary mixed state as the quantum
true data, respectively. The count limit $c_{\mathrm{B}}$ for these
two cases is 500 and 300, respectively. The obtained average $c_{\mathrm{step}}$
for these two types of adversarial learning process is 243 and 170
respectively. \textbf{(b)} The cumulative probability of final state
fidelity $F$. The average fidelities for the pure and the mixed state
are both $98.8\%$. For comparison, the noiseless
numerically simulations of the adversarial learning process are also
performed for 100 times, and their distributions are shown
as solid lines.}
% Comparing the theoretical distributions for 100 and 1000 times, we can get that 100 is enough to reflect the real distribution. 
\label{fig:Fig4} \vspace{-6pt}
 
\end{figure*}

The protocol of our experimental QGAN algorithm is illustrated in
Fig.~\ref{fig:Fig2}(a). The experiment starts with a randomly generated
state $\rho(r_{0},\theta_{0},\varphi_{0})$ by G, a randomly picked
measurement axis $\mathcal{M}(\beta_{0},\gamma_{0})$ by D, and the
quantum true data $\sigma$ from a fixed channel simulator. In each
round of experiment, D plays the adversarial game first with $\rho$ fixed, followed by G's turn with $\mathcal{M}$ fixed. In all experiments, $d$ is obtained by averaging $n=5,000$ repetitive measurements on the true and the fake data respectively. The gradient $\partial{d}/\partial\xi$
for the control parameter $\xi\in\left\{ r,\theta,\varphi,\beta,\gamma\right\} $
is critical for the QGAN. These gradients are approximately obtained
by measuring ${d}$ with respect to $\xi$ and $\xi+\delta$ $\left(\delta\ll1\right)$
and calculating the differential numerically in a classical computer.
According to the principle of gradient descent, the parameters are
updated to maximize $d$ (minimize $d$ with $d>0$) for D's (G's)
turn, as explained in Fig.$\,$\ref{fig:Fig2}(b) (see Supporting Materials for the strategy). Here, the procedure to estimate each gradient
for the relevant parameters is counted as one step, and the total number of steps quantifies the
consumption of time and copies of data. In practical experiments,
the projective detection outcomes follow a binomial statistic, and
show a standard deviation of ${d}$ as $\mathrm{sd}=\sqrt{p_{\rho}\left(1-p_{\rho}\right)/n+p_{\sigma}\left(1-p_{\sigma}\right)/n}$.
When approaching the Nash equilibrium, $p_{\rho}\approx p_{\sigma}\approx\frac{1}{2}$,
then $\mathrm{sd}\approx1/\sqrt{2n}=0.01$. Therefore,
the measurement precision of ${d}$ will limit the convergence
of the game. In our experiments, D's turn ends when the differences
of ${d}$ in the last 3 steps are less than $0.02$. The G's turn
ends when ${d}<R_{j}$ for the $j$th round: $R_{j}=0.055-0.01j$
when $j\leq3$ and $R_{j}=0.02$ when $j>3$. These two adversarial
learning procedures can be repeated many rounds until either the total
count of steps $c_{\mathrm{step}}$ reaches a preset limit $c_{\mathrm{B}}$
or the optimized ${d}$ in D's round is smaller than a preset bound
$d_{\mathrm{B}}$.

Figures~\ref{fig:Fig3}(a-c) show the typical results for the experimental
QGAN with $\sigma=\ket{g}\bra{g}$ of the transmon qubit, the highest purity state, as an
example for the quantum true data. Since a digital quantum channel simulator can generate an arbitrary quantum state~\cite{Hu2018b}, the QGAN experiments by taking an arbitrary mixed state of the transmon as the true data is also studied and the results are depicted in Figs.~\ref{fig:Fig3}(d-f). During the QGAN, the trajectory of control parameters are recorded
{[}Figs.$\,$\ref{fig:Fig3}(b) and (e){]} instead of characterizing
the exact experimental $\rho$ and $\mathcal{M}$. As shown in Figs.$\,$\ref{fig:Fig3}(a)
and (d), the snapshots of the quantum states and measurement axis
at the particular steps indicated by the arrows in Figs.$\,$\ref{fig:Fig3}(b)
and (e) respectively (from left to right in the same order) are plotted
on the Bloch sphere. Here, $\sigma_{\mathrm{I}}$, $\rho_{\mathrm{I}}$
and $\mathcal{M}_{\mathrm{I}}$ are the ideal results derived based
on the calibrated control parameters. As expected, D adaptively adjusts
$\mathcal{M}$ to be parallel with $\rho_{\mathrm{I}}-\sigma_{\mathrm{I}}$, while G learns from the
measurement outcomes to generate quantum data to fool D, and the generated
quantum data gradually converges to the plane that contains the quantum true
data and is perpendicular to $\mathcal{M}$. As a result, ${d}$ oscillates in the D's and G's turns due to the
adversarial process and eventually approaches $0$, which indicates
that ultimately D fails to discriminate $\rho$ from $\sigma$, and
G achieves his goal of replicating the statistics of the quantum true
data.

To characterize the adversarial learning process, the state fidelity
$F(\sigma_{\mathrm{I}},\rho_{\mathrm{I}})=\mathrm{tr}\sqrt{\sqrt{\sigma_{\mathrm{I}}}\rho_{\mathrm{I}}\sqrt{\sigma_{\mathrm{I}}}}$
in the adversarial process is introduced to quantify the indistinguishability
between the true data and the generated data. As plotted in
Figs.$\,$\ref{fig:Fig3}(b) and (e), $F$ approaches 1 after about
220 and 120 steps, respectively. %\red{It takes a larger count of steps to converge when $\sigma$ is pure because in this case the condition $r=1$ (corresponding to its upper bound) would cause oscillations in the optimization process.} 
The final generated quantum state
$\rho_{\mathrm{f}}$ after the adversarial game and the experimental input state $\sigma_{e}$ are measured by state tomography {[}Figs.$\,$\ref{fig:Fig3}(c) and (f){]}, and a fidelity as large as $F=99.1\%$ for both cases
are achieved. Such high fidelities verify that the unique Nash equilibrium,
in which a quantum generator can replicate the statistics of the quantum
true date, can be efficiently achieved in a quantum experimental realization
of GAN. It is worth noting that, although we calibrate the system
parameters to infer the ideal $\sigma_{\mathrm{I}}$ and $\rho_{\mathrm{I}}$
during the adversarial process, it is not necessary for the QGAN.
Our experimental protocol can reach its equilibrium without requiring
the knowledge about the exact data generated by G or the measurement
axis chosen by D, and thus promises a double-blind quantum machine
learning process just as its classical counterparts.

%  and pasting the end point of $\sigma$
%  After adversarial process, we do the state tomography experiment for $\sigma$ and $\rho$ generated using the last parameters and get their state fidelity. The results means that after the adversarial process $\rho$ converges to $\sigma$ yielding a state fidelity of $97.7\%$. The higher fidelity than the ideal state is because that though the the qubit drives is non-ideally, the optimization can get the optimal parameter for the real system. 

By taking the total steps ($c_{\mathrm{step}}$) and the fidelity
of final generated state ($F$) as the figures of merit, the statistics
of our QGAN performance is finally studied with 100 random adversarial
learning processes. We study both cases with the same pure and arbitrary
mixed states as the quantum true data as in Fig.$\,$\ref{fig:Fig3}, but with different random $\rho$ and $\mathcal{M}$ at the beginning, all showing similar behaviors as those in Fig.$\,$\ref{fig:Fig3}. Figure~\ref{fig:Fig4}(a)
plots the cumulative probability of the total steps, i.e. the probability
to finish the QGAN experiment within $c_{\mathrm{step}}$ steps. The
average $c_{\mathrm{step}}$ for these two types of adversarial learning
process are 243 and 170, respectively. Figure~\ref{fig:Fig4}(b)
shows the cumulative probability of state fidelity $F$ with the average
fidelities for both the pure and the mixed quantum data of $98.8\%$. Comparing to the noiseless numerical
simulation results, the experimental average $c_{\mathrm{step}}$
is about twice larger, and the average $F$ is about $1\%$ lower.
These differences are mainly attributed to the decoherence processes of the qubit, the finite measurement precision
of ${d}$, and the non-ideal measured gradients. Further studies about
the effects of the experimental imperfections are provided in Supporting
Materials.

In conclusion, we have demonstrated the feasibility of quantum generative adversarial learning with a superconducting quantum circuit, in which the input data, the generator and the discriminator are all quantum mechanical. Our results show that the generator can indeed learn the patterns of the input quantum data and produce quantum states with high fidelity that are not distinguishable by the discriminator. 
Since our QGAN experiment requires neither a quantum random accessing memory, nor a universal quantum computing device or any fine-tuning parameters (thus robust to experimental imperfections), it carries over to the noisy intermediate-scale quantum (NISQ) devices widely expected to be available in the near future~\cite{Preskill2018}. An experimental demonstration of QGAN with NISQ devices promises to showcase the quantum advantages over classical GAN---a possible approach to realize quantum supremacy~\cite{Preskill2012Quantum,Harrow2017Quantum} with practical applications. Our results may also have far-reaching consequences in solving quantum many-body problems with QGAN, given the recent rapid progresses in related directions \cite{Carleo2016Solving,Deng2017Quantum,Deng2017MachineBN}. In addition, the hybrid quantum-classical architecture demonstrated in this work can be straightforwardly extended to the optimal control~\cite{Li2017b} and self-guided quantum tomography~\cite{Chapman2016}, and we also anticipate their applications in other quantum machine learning/artificial intelligence algorithms.

%and two logical qubits in separate 3D cavities have already been realized \cite{Wang2016,Chou2018,Rosenblum2018}

%\bibliographystyle{nphys}
 \bibliographystyle{Zou}
\bibliography{bibliography}

\vspace{0.2in}

\vbox{}

\noindent \textbf{Acknowledgments}\\
We thank N. Ofek and Y. Liu for valuable suggestions on FPGA programming. LS acknowledges the support from National Key Research and Development Program of China No.2017YFA0304303, National Natural Science Foundation of China Grant No.11474177, and the Thousand Youth Fellowship program in China. LS also thanks R. Vijay and his group for help on the parametric amplifier measurements. CLZ is supported by Anhui Initiative in Quantum
Information Technologies (AHY130000). D.L.D. acknowledges the start-up fund from Tsinghua University. 

\vbox{}

\end{document}